\documentclass[fleqn,twoside]{article}
\usepackage{espcrc2}
\usepackage{epsfig}

\usepackage{graphicx}
\usepackage[figuresright]{rotating}

\newcommand{\AmS}{{\protect\the\textfont2
  A\kern-.1667em\lower.5ex\hbox{M}\kern-.125emS}}

\title{The Strange Quark Mass, $\alpha_s$, and the Chiral Limit 
Electroweak Penguin $K\rightarrow\pi\pi$
Matrix Elements From Hadronic $\tau$ Decay Data}

\author{K. Maltman\address{Dept. Mathematics and Statistics,
York University, 4700 Keele St., Toronto, Canada M3J 1P3;\\ CSSM, 
University of Adelaide, Adelaide, 5005 Australia}
\thanks{Work supported by the Natural Sciences and Engineering Research
Council of Canada}}
\begin{document}

\begin{abstract}
Hadronic $\tau$ decay data provides 
access to the light quark vector (V) and axial vector (A) 
spectral functions.  This makes
possible investigations of the 
dynamics of QCD at intermediate scales and improved
determinations of certain QCD/Standard Model parameters.
We discuss three such applications: (1) the
investigation of the nature of duality violation 
in QCD at intermediate scales (and its relation to the determination
of $\alpha_s$);
(2) the extraction of $m_s$ from flavor-breaking differences of
$ud$ and $us$ data; and (3) the determination of 
the dimension $D=6$ term in the OPE for the flavor $ud$ V-A
correlator difference.  The latter is relevant to the evaluation
of the chiral limit values of the $K\rightarrow\pi\pi$ 
electroweak penguin matrix elements, and hence to 
our understanding of expectations for
$\epsilon^\prime /\epsilon$ in the Standard Model.
\vspace{1pc}
\end{abstract}

\maketitle

\section{INTRODUCTION/BACKGROUND}
As is well known, the ratio, $R_{V/A;ij}$, of the flavor $ij=ud,us$
vector (V) or axial vector (A) current-mediated hadronic 
$\tau$ decay rate to the corresponding
electron decay rate,
\begin{equation} 
R_{V/A;ij}
= {\frac{\Gamma [\tau^- \rightarrow \nu_\tau
\, {\rm hadrons}_{V/A;ij}\, (\gamma)]} 
{\Gamma [\tau^- \rightarrow
\nu_\tau e^- {\bar \nu}_e (\gamma)]}}\ ,
\end{equation}
(with ($\gamma $) denoting extra photons and/or lepton pairs)
can be written as a sum of weighted integrals
over the spin $J=0$ and $1$ hadronic spectral 
functions\cite{braatenetc,pichrev}.
Explicitly,
\begin{eqnarray}
&&{\frac{R_{V/A;ij}}{\left[ 12\pi^2\vert V_{ij}\vert^2 S_{EW}\right]}}
=\int^{1}_0\, dy_\tau \,
\left( 1-y_\tau\right)^2 \nonumber\\
&&\qquad\left[ \left( 1 + 2y_\tau\right) 
\rho_{V/A;ij}^{(0+1)}(s) - 2y_\tau \rho_{V/A;ij}^{(0)}(s) \right]\nonumber \\
&&\qquad\equiv \int_0^{m_\tau^2}ds\, {\frac{dR_{V/A;ij}(s)}{ds}}
\label{taukinspectral}
\end{eqnarray}
where $y_\tau =s/m_\tau^2$, $V_{ij}$ is the 
flavor $ij$ CKM matrix element, $S_{EW}$ is
an electroweak correction, and
$\rho^{(J)}_{V/A;ij}(s)$, with $(J)$ the spin of the hadronic
system, is the spectral function of the corresponding spin $(J)$
part of the flavor $ij$ V/A correlator,
$\Pi^{(J)}_{V/A;ij}(s)$.{\begin{footnote}
{$\rho^{(0)}(s)+\rho^{(1)}(s)\equiv \rho^{(0+1)}(s)$.  This combination,
and also $s\rho^{(0)}(s)$, correspond to scalar correlators,
$\Pi^{(0+1)}_{V/A;ij}\equiv \Pi^{(0)}_{V/A;ij}+\Pi^{(1)}_{V/A;ij}$
and $s\Pi^{(0)}_{V/A;ij}$, having no kinematic singularities.}\end{footnote}}
The $\Pi^{(0,1)}_{V/A;ij}(s)$ are defined by
\begin{eqnarray}
&&i\, \int\, d^4x\, e^{iq\cdot x}\langle 0\vert T\left( J_{V/A}^\mu (x)
J_{V/A}^\nu (0)\right)\vert 0\rangle =\nonumber\\
&&\quad \left( q^\mu q^\nu -q^2g^{\mu\nu}\right)\Pi^{(1)}_{V/A}(q^2)
+q^\mu q^\nu \Pi^{(0)}_{V/A}\ ,
\end{eqnarray}
where $J_{V/A}^\mu$ are the standard V and A currents.
For a given channel, the kinematically-weighted linear 
combination of $(J)=(0+1)$ and $(0)$ spectral functions
appearing in Eq.~(\ref{taukinspectral}) 
can thus be extracted from the corresponding
bin-by-bin experimental decay distribution.

For the V channel, $\rho_{V;ij}^{(0)}$ is proportional to 
$(m_i-m_j)^2$, and hence numerically negligible for $ij=ud$.
For the A channel, $\rho_{A;ij}^{(0)}$ is saturated by the flavor $ij$
Goldstone boson pole in the chiral limit and has
non-Goldstone-boson contributions proportional to $(m_i+m_j)^2$.  Apart
from the $\pi$ pole contribution, $\rho_{A;ud}^{(0)}$ is, thus, also
numerically negligible.  $m_s$ is not sufficiently small that
$\rho_{V;us}^{(0)}$ and the non-$K$-pole contributions to 
$\rho_{A;us}^{(0)}$ can be safely neglected.
The spin separation of the $us$ experimental data is straightforward
for the $K$ and $K^*$ contributions, but not yet available for the data
above the $K^*$.  Flavor $ud$ V/A separation for states consisting
only of pions can be accomplished using G-parity.

Cauchy's theorem, together with analyticity, implies that
correlators $\Pi$ having no kinematic singularies satisfy 
finite energy sum rules (FESR's).  
For weight functions, $w(s)$, analytic in the region 
$\vert s\vert <S$ of the complex-$s$ plane, and any $s_0<S$, 
these have the form
\begin{equation}
\int_0^{s_0}\, ds\, \rho (s)\, w(s)\, =\,
{\frac{-1}{2\pi i}}\oint_{\vert s\vert =s_0}ds\, \Pi (s)\, w(s)\ .
\label{basicfesr}
\end{equation}
For $s_0$ large enough that $\Pi$ on the RHS of Eq.~(\ref{basicfesr})
may be approximated by its OPE, one
obtains a relation between spectral data 
and OPE parameters.  The well-known predictions
for the inclusive flavor $ij=ud,us$ hadronic $\tau$ decay widths in terms of
the parameters occurring in the OPE 
of the corresponding V and A correlators (dominantly
$\alpha_s$)~\cite{braatenetc},
\begin{eqnarray}
&&R_{V/A;ij}= 6\pi i\, S_{EW}\vert V_{ij}\vert^2\, \oint_{|y|=1} dy 
\left( 1- y\right)^2 \nonumber \\
&&\ \ \times\left[ \left( 
1 + 2y\right) \Pi_{V/A;ij}^{(0+1)}(s)- 2 y\Pi_{V/A;ij}^{(0)}(s) \right] 
\label{kinematicfesr}\end{eqnarray}
are examples of such a relation.
The modified version of this relation obtained by multiplying
the integrands in both the spectral and OPE representations
of $R_{V/A;ij}$ by the factor $(1-y_\tau )^k y_\tau^m$
is called ``the $(k,m)$ spectral weight sum rule''~\cite{pichled92}.
Such sum rules have the advantage that a spin separation of
the $ij=us$ spectral data is not required in order to evaluate
their spectral sides.

The spectral integrals required for various V and A FESR's 
can be evaluated, using hadronic $\tau$ decay data, for
any $s_0\leq m_\tau^2$.  So long as there exists a window
of $s_0$ values below $m_\tau^2$ for which the breakdown
of the OPE representation of the relevant
correlator (``duality violation'') is 
negligible, the values of basic QCD parameters,
such as $\alpha_s$ and $m_s$, appearing on the OPE side
of the FESR relation can be determined using spectral integral data.  
The existence of such a window of $s_0$ values is, of course,
crucial to the reliability of such a determination.  
We will concentrate on three analyses of this type below.

The rest of the paper is organized as follows.
In Section 2 we discuss the issue of duality violation, 
and the closely related issue of the determination of $\alpha_s$, 
using the very precise $ud$ V+A spectral data.
In Section 3, we discuss some complications,
and the current status, of attempts to extract $m_s$ from the 
flavor-breaking difference of $ud$ and $us$ V+A correlators.
In Section 4, we discuss some recent work on the determination
of the dimension $6$ term in the OPE of the $ud$ V-A correlator.
This quantity bears a special relation to the chiral limit values of the 
$K\rightarrow\pi\pi$ electroweak penguin (EWP) operator
matrix elements, and hence to expectations for the value of
$\epsilon^\prime /\epsilon$ in the Standard Model.
Finally, in Section 5, we consider what improvements in
these determinations may be possible in the near future.

\section{DUALITY VIOLATION AND THE EXTRACTION OF $\alpha_s$}
For very large $s_0$, the OPE is expected to provide
a reliable representation of hadronic correlators over the
entire circle $\vert s\vert =s_0$ in the complex $s$-plane.
For such $s_0$, the corresponding spectral function, $\rho (s_0)$,
will also be well-represented by its OPE form.  This is conventionally
referred to as the regime of the validity of ``local duality'' (LD).
As one moves to lower $s_0$, the arguments of 
Poggio, Quinn and Weinberg~\cite{pqw} suggest that the OPE
will break down first for those $s$ on $\vert s\vert =s_0$
near the timelike real axis. We thus expect there to exist 
a regime of ``intermediate'' scales, $s_0$, for which the OPE, while 
not reliable near the timelike real axis, continues to be reliable 
over most of the rest of the circle $\vert s\vert =s_0$.  
We will refer to the scales for which this is true 
as the regime of the validity of ``semi-local duality'' (SLD).
Eventually, for sufficiently small $s_0$, we expect the OPE
to become unreliable over the whole of the circle $\vert s\vert =s_0$.
Were all $s_0<m_\tau^2$ to lie in this final regime, 
it would be impossible to use hadronic $\tau$
decay data to determine parameters appearing in the OPE; it is
therefore important to verify that some of the scales kinematically 
accessible in $\tau$ decay lie, at the very least, in the 
region of validity of SLD.

The flavor $ud$ V, A and V+A correlators provide an excellent
laboratory for studying the nature of duality violation.  The
reason is that, for scales $\sim 2-3\ {\rm GeV}^2$, the OPE
for these correlators is completely dominated
by the $D=0$ term, which is known to $O(\alpha_s^3)$~\cite{alphas3}.
Since $\alpha_s$ is measured with good accuracy at the $Z$ scale
and the QCD $\beta$-function is known to $4$-loop order~\cite{beta4},
so that $\alpha_s$ can be run down reliably to scales below 
$2\ {\rm GeV}^2$, 
the OPE in these cases is known with good accuracy down to scales
significantly below $m_\tau^2$.  One can, therefore, compare the
spectral and OPE integrals for various FSER weight choices and
determine the extent to which the OPE provides a good representation
of the spectral integral data.  

The non-perturbative terms of the OPE 
relevant to this comparison are~\cite{bnp}
\begin{equation}
\left[\Pi^{(0+1)}_{V/A;ud}\right]_{D=4}={\frac{\left(
1-{\frac{11}{18}} a\right)}{12 Q^4}}\langle aGG\rangle
+\cdots\label{udd4}\end{equation}
\begin{equation}
\left[\Pi^{(0+1)}_{V/A;ud}\right]_{D=6}={\frac{\pi\rho
\alpha_s\langle \bar{q}q\rangle^2}{Q^6}}
\left({\frac{(\mp 288 +64)} {81}}\right) 
\label{udd6}
\end{equation}
where $a\equiv a(Q^2)= \alpha_s(Q^2) /\pi$, $\rho\equiv\rho_{VSA}$ 
expresses the 
deviation of the four-quark condensates from their vacuum
saturation approximation (VSA) values, $+\cdots$
in Eq.~(\ref{udd4}) stands for numerically small terms involving
the quark condensates and fourth powers of the light
quark masses, and, in Eq.~(\ref{udd6}),
the upper (lower) sign refers to the V (A) case.  
The dominant $D=0$ contribution may be
expressed in terms of the Adler function
\begin{eqnarray}
D_{V/A}(Q^2)&&\equiv -Q^2\left[
{\frac{d\Pi^{(0+1)}_{V/A}}{Q^2}}\right]_{D=0}\nonumber\\
&&= {\frac{1}{4\pi^2}}\, \sum K_n\, a(Q^2)^n\ ,\label{adler}
\end{eqnarray}
where, in the $\overline{MS}$ scheme, $K_0=K_1=1$, $K_2=1.63982$,
$K_3=6.37101$~\cite{alphas3}, and the higher $K_n$ are unknown.

To be specific, we take as OPE input 
(1) $\alpha_s(m_\tau )=0.334\pm 0.022$~\cite{ALEPHud,ALEPH99},
(2) $\langle {\frac{\alpha_s}{\pi}}G^2\rangle = (0.009\pm .015)\ 
{\rm GeV}^4$,{\begin{footnote}{This value encompasses 
the ranges obtained in three different analyses,
$0.021\pm 0.03\ {\rm GeV}^4$~\cite{narisonaGG}, 
$0.006\pm 0.012\ {\rm GeV}^4$~\cite{giz01} and
$0.009\pm 0.007\ {\rm GeV}^4$~\cite{iz02}.}\end{footnote}} 
(3) $\rho_{VSA}=1\pm 5${\begin{footnote}{The VSA-violating
parameter, $\rho_{VSA}$, for the V-A correlator 
is known to be $\sim 1.6$~\cite{cdgm02}.  Significant cancellation,
however, occurs in the VSA expression for the V+A sum; the expanded
error, $\pm 5$, on the VSA value $\rho_{VSA}=1$ is meant to deal
with this in a conservative manner.}\end{footnote}} and use
$K_4=25\pm 50$ as a means of 
estimating the $D=0$ truncation error~\cite{pichrev}.  
In the V+A case, the size of the VSA $D=6$ contribution 
is quite small, so the conservative error on $\rho_{VSA}$
has little impact on the full theory errors, which are dominated by 
the uncertainty in $\alpha_s(m_\tau )$ and the 
estimated truncation error.  Possible $D=8$ and higher terms
are neglected.  Errors associated with the uncertainty in each of
the inputs are combined in quadrature to obtain the total theory
error.  Spectral integrals are evaluated using the
ALEPH data~\cite{ALEPHud,ALEPH99}, which produces the smallest 
errors{\begin{footnote}{See also Refs.~\cite{OPALlight} for
the OPAL data on the $ud$ V and A spectral functions.  The
spectral integrals using the OPAL data are in good agreement
with those using ALEPH data.}\end{footnote}}.  The corresponding
errors are computed using the ALEPH covariance matrix.
The final errors on the ratios of OPE to data integrals,
which provide a measure of duality violation,
are obtained by combining theory and experimental errors in quadrature.

Making the OPE/data comparison first for the V channel, we discover that
$s_0<m_\tau^2$~\cite{kmfesr} is not a region of the validity of
LD for the $ud$ V correlator.  Indeed, the OPE and spectral
integrals for FESR's
based on the weights $w(s)=s^k$ (which do not suppress contributions
from the vicinity of the timelike real axis) are in rather poor
agreement.  This is illustrated in Figure 1, which displays
the OPE-to-spectral-integral-ratio (which should be $1$
if LD is valid) as a function of $s_0$,
for $k=0,\cdots ,3$.  
Also shown are the corresponding results for the $ud$
V+A combination.  One can see that the level of duality
violation in the V+A correlator is much smaller than that
in the V and A correlators separately.

\begin{figure*}
\unitlength1cm
\caption{Ratios of OPE to data integrals as a function of
$s_0$ for the flavor $ud$ V (open circles) and V+A correlators
(solid circles).  The top left, top right, bottom left and
bottom right figures correspond to $w(s)=1$, $w(s)=s$,
$w(s)=s^2$ and $w(s)=s^3$, respectively.}
\begin{minipage}[t]{8.0cm}
\begin{picture}(7.9,7.9)
\epsfig{figure=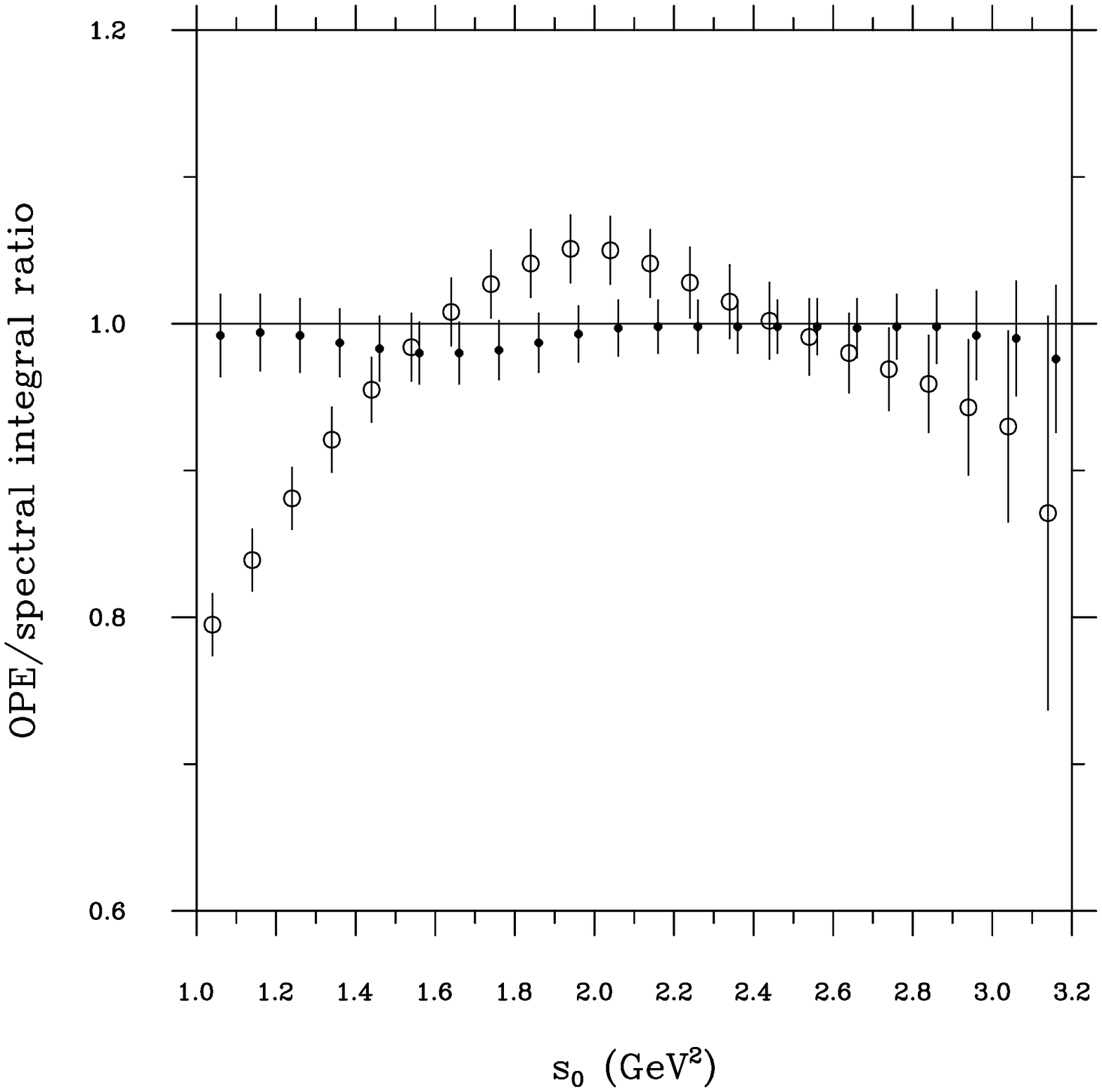,height=7.8cm,width=7.8cm}
\end{picture}
\end{minipage}
\hfill
\begin{minipage}[t]{8.0cm}
\begin{picture}(7.9,7.9)
\epsfig{figure=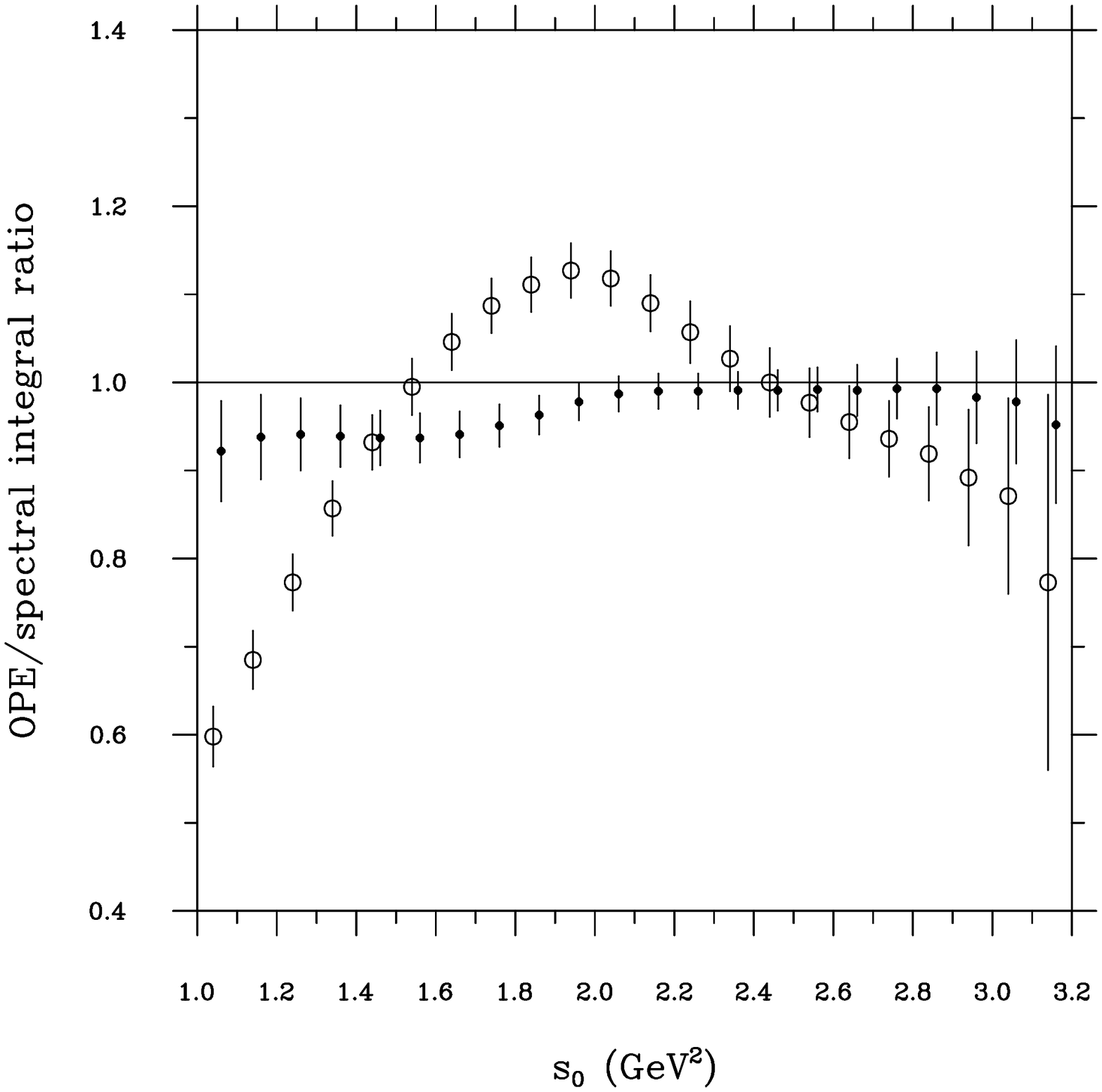,height=7.8cm,width=7.8cm}
\end{picture}
\end{minipage}
\vskip .15in\noindent
\begin{minipage}[t]{8.0cm}
\begin{picture}(7.9,7.9)
\epsfig{figure=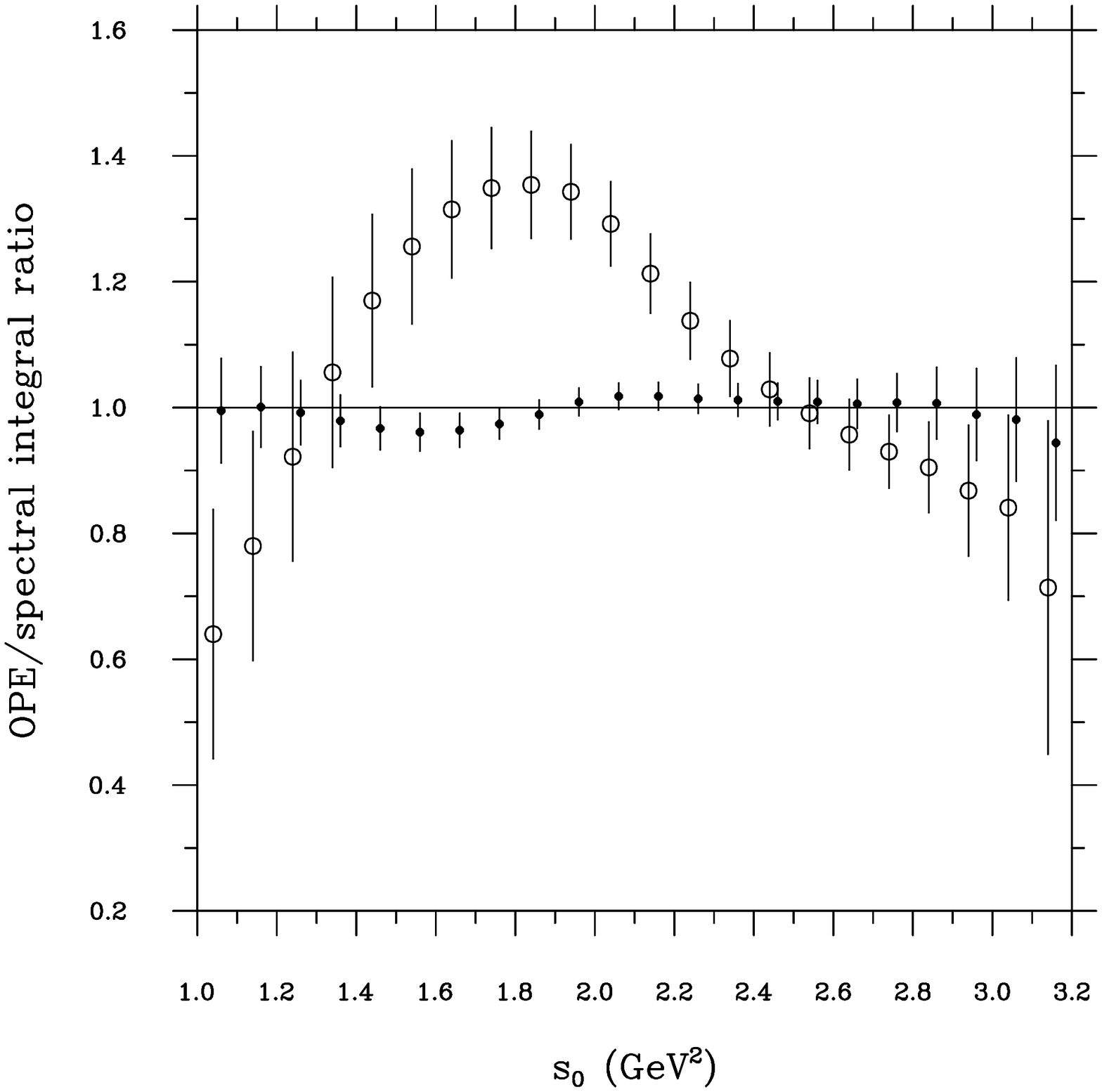,height=7.8cm,width=7.8cm}
\end{picture}
\end{minipage}
\hfill
\begin{minipage}[t]{8.0cm}
\begin{picture}(7.9,7.9)
\epsfig{figure=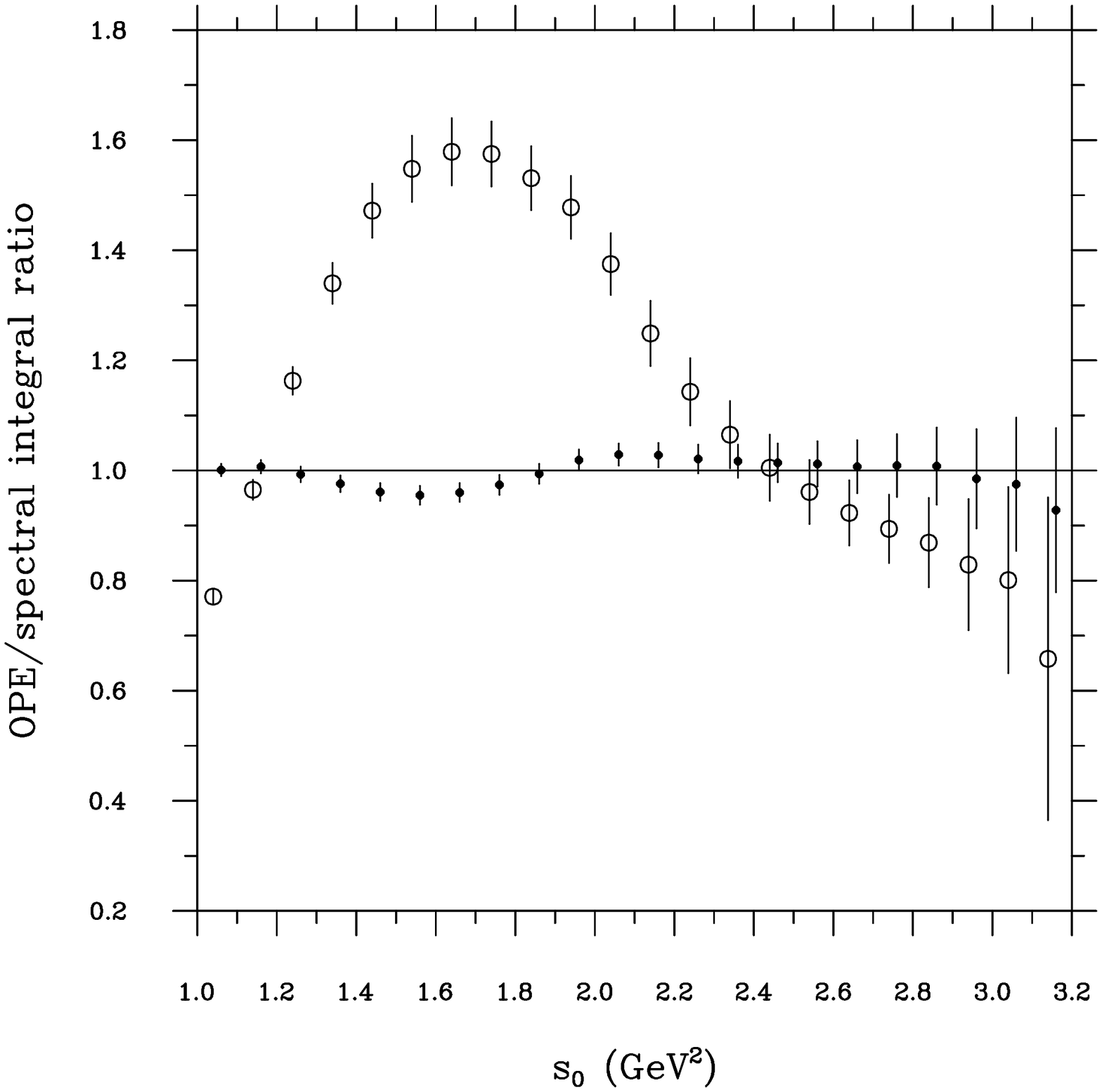,height=7.8cm,width=7.8cm}
\end{picture}
\end{minipage}
\label{Figure:1}\end{figure*}

The strong duality violations in the $s^k$-weighted
FESR's for the $ud$ V correlator are not a general feature 
of FESR's involving this correlator.  Indeed, 
the FESR for $R_{V;ud}$, Eq.~(\ref{kinematicfesr}),
(whose ``kinematic'' weight, 
$w_\tau \left( y_\tau\right)=\left( 1-y_\tau\right)^2\left( 1+2y_\tau\right)$, 
is a linear combination of the four $s^k$-weighted FESR's
shown in Figure 1) is known
to be in good agreement with experiment~\cite{ALEPHud,ALEPH99}.  
The difference has to do with the form of the weight: because
the point $s=m_\tau^2$, where the circle 
$\vert s\vert =m_\tau^2$ crosses the timelike real axis,
lies at the edge of hadronic phase space,
$w_\tau$ has a (double) zero at $s=m_\tau^2$.  This zero
suppresses OPE contributions from the part of the integration region 
near the timelike real axis where use of the OPE 
is expected to be potentially most problematic.  It is
presumably this feature of $w_\tau$
which is responsible for the success of the $R_{V;ud}$ FESR.

If, as this argument suggests, the success of the $R_{V;ud}$ FESR 
is a reflection of the localization of the breakdown of the OPE
to the vicinity of the timelike real axis for scales $s_0\simeq m_\tau^2$,
then other FESR's with suppressions of 
OPE contributions from this region should 
also be well satisfied.  This hypothesis 
was tested in Ref.~\cite{kmfesr}.  FESR's for the V channel 
were considered for a range of 
$s_0$ different from $m_\tau^2$,
and a range of alternate weights, $w(s)$, still satisfying 
$w(s=s_0)=0${\begin{footnote}{We refer 
to such weights as ``pinched'' weights, and the FESR's based on
them as ``pinched FESR's'', or pFESR's.  The natural variable for
use in such pFESR's is $y\equiv s/s_0$.}\end{footnote}}.
The resulting pFESR's were all found to be well satisfied for
all $s_0>2\ {\rm GeV}^2$.  This suppression of duality violations,
which is a generic feature of weights of the forms
\begin{eqnarray}
&&w(y)=(1-y)^2(1+Ay) 
\label{doublepinch}\\
&&w(y)=(1-y)(1+Ay) \ ,
\label{singlepinch}
\end{eqnarray}
having either a single or double zero at $s=s_0$
(where $A$ is a free parameter) can, in favorable
cases (including that of the kinematic weight),
extend to much smaller $s_0$.  This is illustrated for
the pFESR's based on the weights $w(y)=(1-y)^2(1+2y)$ and
$w(y)=1-y$ in the left and right panels of Fig. 2,
respectively.
Also shown for comparison in each panel are the results for the 
corresponding $s^k$-weighted (unpinched) FESR's, where
$k$ is the degree of the pinched weight in question.
The $w(y)=1-y$ pFESR
receives no contributions from the $D=6$ OPE term and
hence has reduced OPE errors at low $s_0$.  The absence
of duality violation in this case extends to very low $s_0$,
despite the fact that the weight has only a single zero at $s=s_0$.

\begin{figure*}
\unitlength1cm
\caption{Ratios of OPE to data integrals as a function of
$s_0$ for the pinched and unpinched FESR's
involving the $ud$ V correlator.
The left figure shows the results for
$w(y)=(1-y)^2(1+2y)$ (solid circles) and $w(s)=s^3$ (open circles),
the right figure the results for
$w(y)=1-y$ and $w(s)=s$.}
\begin{minipage}[t]{7.5cm}
\begin{picture}(7.4,7.4)
\epsfig{figure=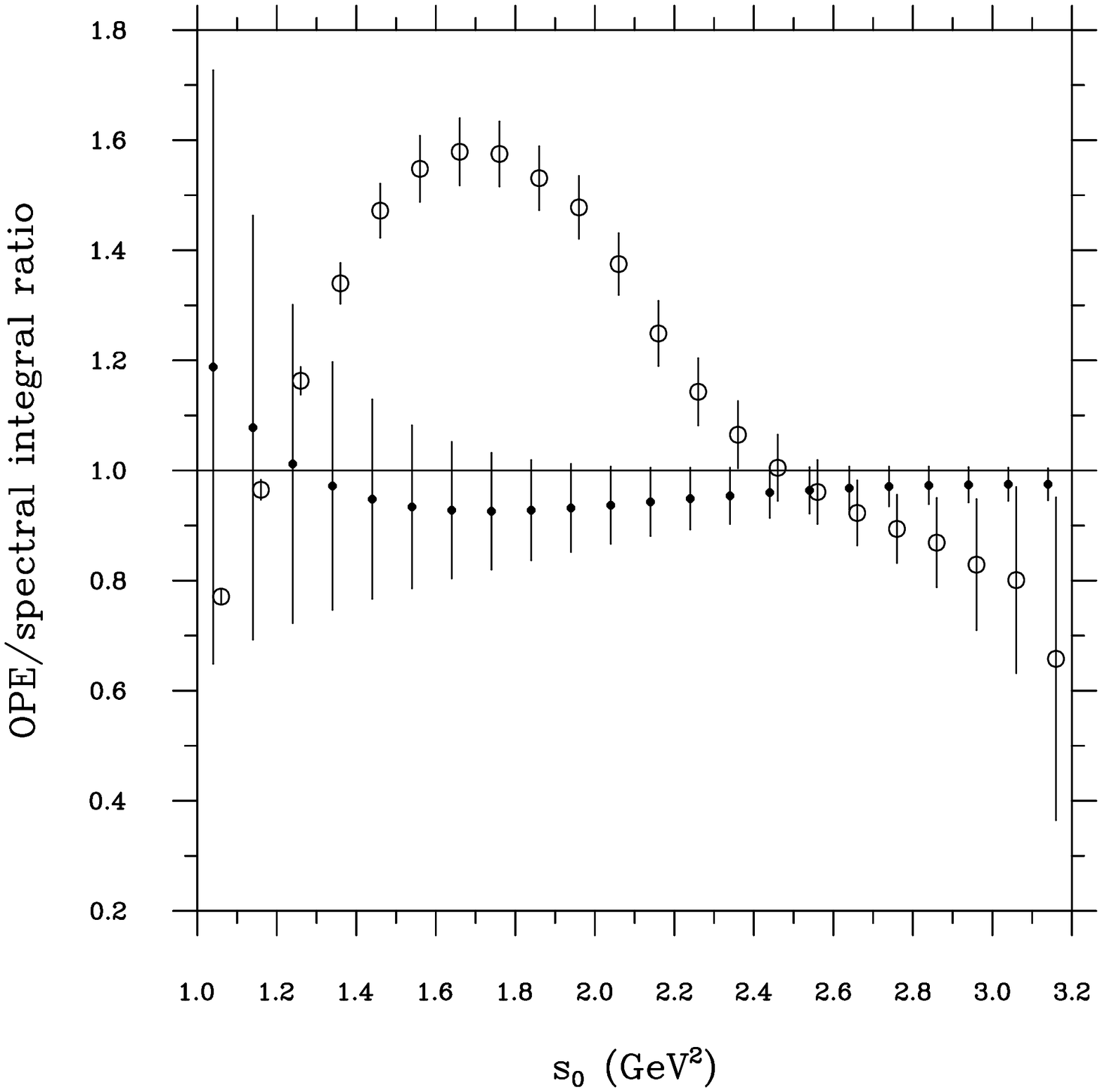,height=7.3cm,width=7.3cm}
\end{picture}
\end{minipage}
\hfill
\begin{minipage}[t]{7.5cm}
\begin{picture}(7.4,7.4)
\epsfig{figure=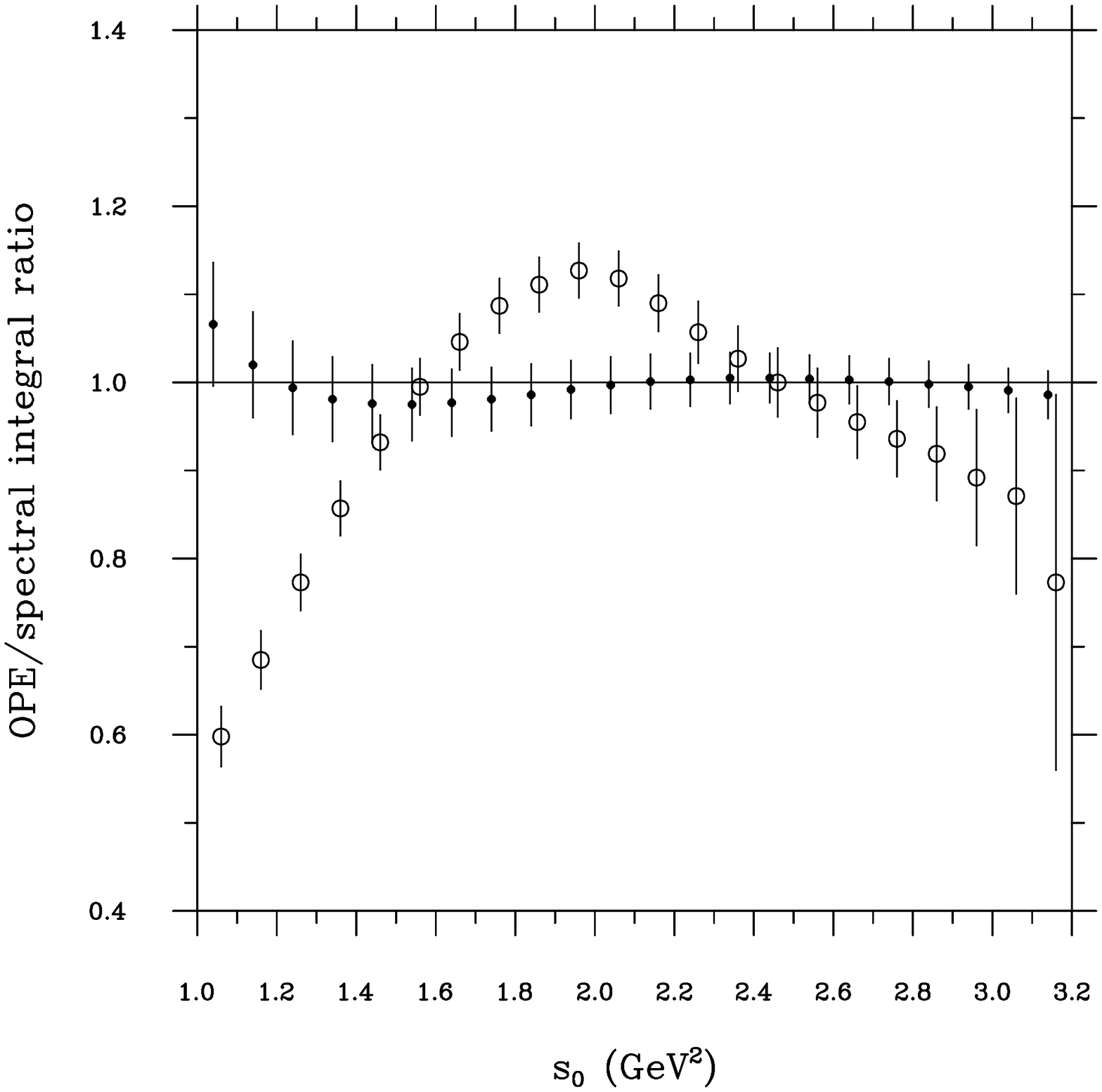,height=7.3cm,width=7.3cm}
\end{picture}
\end{minipage}
\label{Figure:2}\end{figure*}

The results above provide a specific illustration of
the dangers of attempting to extract OPE parameters
from a sum rule analysis at intermediate scales when
duality violations are present.  Were one to take a typical
scale $s_0<m_\tau^2$ and ``extract'' $\alpha_s(s_0)$
by matching the OPE and spectral sides of one of the
$s^k$-weighted FESR's, one would find a value which,
when run up to higher scales, failed to agree with
that obtained in direct measurements, for example, at
the Z scale.  In this case we do not actually need to know
the high-scale value beforehand in order to expose the
presence of duality violation:  the $s_0$-dependence
of the $\alpha(s_0)$ extracted at different $s_0$, but using
the same $s^k$ weight, can been seen to be in very
poor agreement with that predicted by 4-loop running
in QCD.  Similar ``internal'' tests can be performed
for any FESR.  Passing such a test is of course a necessary, but
not sufficient, condition for demonstrating the absence
of duality violations in the analysis in question.

Let us now turn to $\alpha_s$.
In order to reduce the difficult-to-quantify theoretical systematic
error associated with possible residual duality violations,
it is useful to work with correlators which display a reduced
level of duality violation to begin with.  The results of
Figure 1 show that the $ud$ V+A combination is favored in
this regard{\begin{footnote}{This 
combination also has reduced $D=6$ contributions
on the OPE side, and reduced errors for $s>2\ {\rm GeV}^2$
on the experimental side.}\end{footnote}}.
Similarly, pFESR's are favored over their
unpinched analogues at intermediate scales such as those
forced on us by $\tau$ decay kinematics.  Since there is still
significant uncertainty in the value of the gluon condensate,
which dominates the non-perturbative corrections on the OPE
side, it is useful to work with pFESR weights which strongly
suppress the $D=4$ contribution (Eq.~(\ref{doublepinch}) with
$A=2$ and Eq.~(\ref{singlepinch}) with $A=1$).

Bearing in mind the necessity of using the data not only to
extract $\alpha_s$, but also to verify that duality violations
are not obviously present, the following procedure seems
optimally conservative.  We first extract $\alpha_s$ using the 
$w(y)=(1-y)^2(1+2y)$ pFESR at $s_0=m_\tau^2$.  
A weight with a double zero and the highest possible value for $s_0$ are
chosen to increase the likelihood that duality violation will be negligible.
Neglect of the $D=8$ OPE contribution (which is
not suppressed by any factor of $\alpha_s$ for
this weight choice, but which scales as $1/s_0^3$)
is also most reliable for the highest possible $s_0$.
Using the contour improved peturbation theory (CIPT)
scheme~\cite{cipt,pichled92}
for the $D=0$ contribution, the result is
\begin{equation}
\alpha_s(m_\tau )=0.345\pm 0.026
\label{alphafit}\end{equation}
where experimental and theoretical errors have been combined in
quadrature.  If instead of CIPT one employs 
fixed order perturbation theory (FOPT),
the central value is reduced to 
$0.326${\begin{footnote}{FOPT involves an expansion
in $\alpha_s(s_0)$, CIPT a summing of logs point-by-point
around the integration contour by the scale
choice $\mu^2 =Q^2$.}\end{footnote}}.
The method, and results, are similar to those
of earlier analyses~\cite{np93,CLEOlight,ALEPHud,ALEPH99,OPALlight}.

Having extracted $\alpha_s$ using only one pFESR, and one value of
$s_0$, we can then use other pFESR's and other $s_0$ values to
check for possible duality
violation and/or the presence of neglected higher dimension 
contributions{\begin{footnote}{Contributions to the OPE side
from terms of dimension $D=2k+2$ scale as $1/s_0^k$, allowing
one to distinguish contamination by an operator neglected
when it should not have been, at least if one considers the
data integrals over a sufficiently large window of $s_0$ 
values.}\end{footnote}}.  Here we 
compare the OPE predictions (obtained using the
value of $\alpha_s$ just extracted) to the corresponding data integrals for
(1) $w(y)=(1-y)^2(1+2y)$ (at $s_0<m_\tau^2$) and
(2) $w(y)=(1-y)(1+y)$ (also as a function of $s_0$). 
The $s_0$ dependence of the OPE prediction depends crucially
on the form of the 4-loop running of $\alpha_s$.
The results of the tests for possible duality
violation are shown in Figure 3.  There is no evidence of
such violations, within experimental errors, even down to rather
low scales.  While this does not prove that they are absent,
it does gives us additional confidence in the fit value,
Eq.~(\ref{alphafit}).

\begin{figure}[h]
\unitlength1cm
\caption{The OPE and spectral integral sides of two pFESR's
for the $ud$ V+A correlator. The 
value $\alpha_s(m_\tau )=0.345$ obtained by the fitting
procedure described in the text is used as input
on the OPE sides.  The upper curve
corresponds to $w(y)=(1-y)(1+y)$, the lower curve to
$w(y)=(1-y)^2(1+2y)$.}
\begin{minipage}[t]{7.8cm}
\begin{picture}(7.6,7.6)
\epsfig{figure=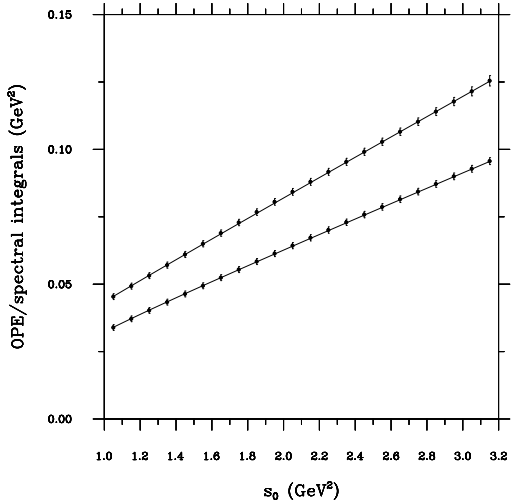,height=7.4cm,width=7.4cm}
\end{picture}
\end{minipage}
\label{Figure:3}\end{figure}

We conclude this section with a comment on
the use of higher degree spectral weights for analyses of the
type described above.  Weights with degrees $k>2$ in principle
involve OPE contributions with $D=8,\cdots , 2k+2$.  Usually
these contributions are assumed to be negligible, at least
for $s_0=m_\tau^2$.  Although the values of the relevant
condensate combinations are not known for the $ud$ V+A
correlator, the analogous V-A combinations have been extracted
in Ref.~\cite{cdgm02}.  If we take these as representative
of the scale of higher $D$ contributions for the V+A
correlator, we can make a rough estimate of the expected
size of higher $D$ contributions, and hence determine the
likelihood that neglect of such contributions is safe.
For the $(0,0)$ spectral weight, $w(y)=(1-y)^2(1+2y)$,
this yields an estimate of $\sim 0.3\%$ of the OPE total
for $s_0=m_\tau^2$, making neglect of such contributions
quite safe.  For the $(1,0)$ spectral weight,
$w(y)=(1-y)^3(1+2y)$, the
estimated $D=8,10$ contributions,
at $s_0=m_\tau^2$, are both $\sim 1.5\%$ of
the OPE total, while for the $(2,0)$ spectral weight,
$w(y)=(1-y)^4(1+2y)$,
the estimated $s_0=m_\tau^2$ $D=8,10,12$ contributions are
$\sim 3\%,\ 6\%$ and $2\%$ of the OPE total, respectively.
Because of the growth in the size of the relevant polynomial
coefficients, the neglect of unknown higher $D$ contributions
thus becomes progressively less safe as one goes to higher
spectral weights.

\section{THE STRANGE QUARK MASS FROM FLAVOR BREAKING IN HADRONIC $\tau$
DECAY}
The $D=0$ (mass-independent) part of the perturbative
contribution to the flavor-breaking difference
$\Delta\Pi_{V/A}^{(J)}(Q^2)\equiv
\left[\Pi_{V/A;ud}^{(J)} - \Pi_{V/A;us}^{(J)}\right] (Q^2)$ vanishes
in the $SU(3)$ flavor limit.
The formally leading (in dimension) 
term in the OPE of $\Delta\Pi_{V/A}^{(J)}$
is therefore the $D=2$ term resulting from flavor breaking in the 
$m_q^2$-dependent perturbative contributions.  
To the extent that $\hat{m}\equiv (m_u+m_d)/2 <<m_s$~\cite{leutwylerqmasses},
this contribution is proportional to $m_s^2$.
The basic idea of using hadronic $\tau$ decay data to determine
$m_s$ is then to construct flavor-breaking differences
of $ud$ and $us$ spectral integrals and equate these,
using the basic FESR relation, to equivalent OPE contour integrals
involving the unknown parameter $m_s$.

A simple way to construct such flavor-breaking differences is
to note that, from Eq.~(\ref{taukinspectral}), the 
$ud$ decay distribution, rescaled by $1/\vert V_{ud}\vert^2$,
and the $us$ decay distribution, rescaled by $1/\vert V_{us}\vert^2$, 
become equal in the $SU(3)_F$ limit.  Weighted integrals of the
difference of the rescaled spectral functions
then produce flavor-breaking observables of the
type amenable to a pFESR extraction of $m_s$.

Such differences can, in principle, be constructed separately
for V, A and V+A combinations, and also separately for $(J)=(0+1)$, $(0)$.
Since the V/A separation has not been performed for the $us$
spectrum, and even for the $ud$ spectrum
the V+A spectral distribution is better
determined than are the separate V and A 
distributions, it is preferable to work with the V+A combinations for both 
$ud$ and $us$.  Bearing in mind that a
$J=0/J=1$ spin separation is not currently available above
$\sim 1\ {\rm GeV}$ in the $us$ channel (where the
$J=0$ spectral contributions to the experimental decay
distributions may not be totally negligible) the natural
choices for flavor-breaking observables are those
constructable from the kinematically-weighted combinations,
$\left( 1-y_\tau\right)^2 \left[
\left(1+2y_\tau\right) \rho_{V+A;ij}^{(0+1)}(s)
- 2y_\tau \rho_{V+A;ij}^{(0)}(s)\right]$,
which are determined directly from the experimental $ij=ud$ and
$us$ decay distributions.  Most analyses
in the literature have employed the $(k,n)$ spectral weight
versions of this construction,
\begin{equation}
\delta R^{(k,n)}\equiv {\frac{R^{(k,n)}_{V+A;ud}}{\vert V_{ud}\vert^2}}
- {\frac{R^{(k,n)}_{V+A;us}}{\vert V_{us}\vert^2}}\ ,
\label{msspecweight}\end{equation}
where
\begin{equation}
R^{(k,n)}_{V+A;ij}\equiv \int_0^{m_\tau^2}\, ds\, 
\left( 1-y_\tau\right)^k y_\tau^n {\frac{d R_{V+A;ij}}{ds}}\ .
\label{defnrkn}\end{equation}

The approach to determining $m_s$ just outlined, though straightforward
in principle, turns out to have non-trivial complications.

On the experimental side, the first complication is 
the limited accuracy of the current determination
of the $us$ spectral distribution (known to $\sim 6-8\%$ in
the $K^*$ region, and to $\sim 20-30\%$ above 
$1\ {\rm GeV}^2$~\cite{ALEPH99}).  
This situation will improve dramatically as 
analyses of the $\tau$ decay data from the B factory experiments begin to 
come online.  The second complication arises from the rather
close cancellation between $ud$ and $us$ contributions occuring on
the spectral sides of the flavor-breaking pFESR's 
employed in the literature (typically to the $\sim 10\%$ or less level).
Such close cancellation makes the $ud$-$us$ spectral difference
sensitive to both uncertainties in $\vert V_{us}\vert^2$
and (apparently) small changes in the measured $us$
branching fraction.  For example, with $ud$-$us$ cancellation
to the $10\%$ ($5\%$) level, the $\sim 2.5\%$ uncertainty
in the value of $\vert V_{us}\vert^2$~\cite{pdg2002}
translates into a $25\%$ ($50\%)$ uncertainty in 
the value of the integrated
$ud$-$us$ spectral difference.  The small differences beween
the preliminary ALEPH determination 
$R_{us}\equiv R_{V+A;us}=0.155\pm 0.006$~\cite{ALEPHprelim},
the final published version, $R_{us}=0.161\pm 0.007$~\cite{ALEPH99},
and Davier's update at Tau'2000, 
$R_{us}=0.163\pm 0.006$~\cite{DHPPC00,CLEOnew,OPALnew},
also produce significant shifts in the
central value of $m_s$ extracted in a given pFESR analysis
for the same reason.  

Table 1 displays the
central values of $m_s$ obtained in a number of independent analyses
reported in the literature, all nominally based on the ``same'' (ALEPH) 
$us$ data.  Also shown are the central values employed for the input 
quantities $R_{us}$ and $\vert V_{us}\vert$.
The entries are labelled by their analysis type.
Table 2 shows the same results converted to common input,
and also to a common $D=2$ OPE truncation scheme, for two
choices of the CKM input: (1) the central values of 
the three-family unitarity-constrained PDG2002 fit, 
$\vert V_{ud}\vert = 0.9734$ and  $\vert V_{us}\vert = 0.2225$ (CKMU) and 
(2) the best fit independent PDG2002 central fit values, 
$\vert V_{ud}\vert = 0.9749$ and  $\vert V_{us}\vert = 0.2196$ (CKMN).  
The sensitivity to $R_{us}$ and $\vert V_{us}\vert$, as well as the
good consistency between the different analyses {\it when the
same input is employed}, is evident from the tables{\begin{footnote}{For
further details on the variations in input among
the different analyses, the impact 
of these variations on the extracted values of $m_s$,
and the conversion to common
input and truncation scheme, see Ref.~\cite{gmdpf00}.}\end{footnote}}.

\begin{table*}
\begin{center}
\caption{Central values for $m_s(2\ {\rm GeV)}$ (in MeV) in the
$\overline{MS}$ scheme from various analyses.
$(k,0)$ labels analyses which combine the
$(0,0)$, $(1,0)$ and $(2,0)$ spectral weights.  The 
1999 ALEPH analysis, labelled by $(k,n)$, is a combined analysis
using the $(0,0)$, $(1,0)$, $(2,0)$, $(1,1)$, $(1,2)$ spectral weights.
The $w_{20}$ label for KM00 denotes a non-spectral weight,
the details of which may be found in Ref.~\cite{km00}; the
analysis in this case is of the non-inclusive $(0+1)$ type.
The various analyses also display some differences in their
truncation procedures, especially in the case of ALEPH99,
and some differences in their treatments of $D=4$
contributions.  Small differences in the central values
employed for $\vert V_{ud}\vert$ also exist.
Details may be found in Ref.~\cite{gmdpf00}.}
\label{table:1}
\newcommand{\m}{\hphantom{$-$}}
\newcommand{\cc}[1]{\multicolumn{1}{c}{#1}}
\renewcommand{\tabcolsep}{2pc} 
\renewcommand{\arraystretch}{1.2} 
\vskip .15in\noindent
\begin{tabular}{@{}llllc}
\hline
Reference           &Type &$R_{us}$ &$\vert V_{us}\vert$ &
$m_s(2\ {\rm GeV})$ [MeV]\\
\hline
CKP98~\cite{ckp98}&$(0,0)$&$0.155$&$0.2213$&$145$\\
ALEPH99~\cite{ALEPH99}&$(k,n)$&$0.161$&$0.2218$&$144$\\
PP99~\cite{pp99}&$(k,0)$&$0.161$&$0.2218$&$114$\\
KKP00~\cite{kkp00}&$(0,0)$&$0.161$&$0.2218$&$125$\\
KM00~\cite{km00}&$w_{20}$&$0.161$&$0.2196$&$115$\\
DHPPC00~\cite{DHPPC00}&$(k,0)$&$0.163$&$0.2225$&$108$\\
\hline
\end{tabular}\\[2pt]
\end{center}
\end{table*}

\begin{table*}
\begin{center}
\caption{Central values for $m_s(2\ {\rm GeV)}$ (in MeV) 
obtained by updating analyses in the literature to common CKM input,
the most recent value of $R_{us}$ and,
where possible, a common truncation scheme.
Possible $D=8$ and higher terms have been neglected throughout.
Since DHPPC00 represents a combined update of both ALEPH99
and PP99, we display only the update of DHPPC00 in this table.
KKP00 employs an expansion in an effective coupling which
is different from that used in the other analyses.
KM00 is a $(0+1)$ analysis, whereas the others are inclusive.
In Ref.~\cite{newpichALEPH} a $k$-dependent truncation
scheme has been used for the $(k,0)$ spectral weights;
this scheme has been retained in converting to CKMU and CKMN
input.  A discussion of the issues underlying this choice of
truncation scheme is given below.
For further details see Ref.~\cite{gmdpf00}.}
\label{table:2}
\newcommand{\m}{\hphantom{$-$}}
\newcommand{\cc}[1]{\multicolumn{1}{c}{#1}}
\renewcommand{\tabcolsep}{2pc} 
\renewcommand{\arraystretch}{1.2} 
\vskip .15in\noindent
\begin{tabular}{@{}llll}
\hline
Reference           &Type &CKMU input
&CKMN input\\
\hline
CKP98~\cite{ckp98}&$(0,0)$&$116\pm 31$&$99\pm 34$\\
KKP00~\cite{kkp00}&$(0,0)$&$120\pm 28$&$106\pm 32$\\
KM00~\cite{km00}&$w_{20}$&$110\pm 16$&$100\pm 18$\\
DHPPC00~\cite{DHPPC00}&$(0,0)$&$124\pm 32$&$106\pm 37$\\
&$(1,0)$&$113\pm 32$&$102\pm 21$\\
&$(2,0)$&$\ 99\pm 21$&$\ 91\pm 21$\\
CDGHPP01~\cite{newpichALEPH}&$(0,0)$&$126\pm 31$&$107\pm 35$\\
&$(1,0)$&$116\pm 19$&$106\pm 19$\\
&$(2,0)$&$113\pm 22$&$103\pm 22$\\
\hline
\end{tabular}\\[2pt]
\end{center}
\end{table*}

On the OPE side, the major complication has to do with the bad
behavior of the various weighted integrals 
of the $D=2$ part of the longitudinal ($(J)=(0)$) contribution,
\begin{equation}
\oint_{\vert y\vert =1}ds\, w(y)\, \left[\Delta\Pi^{(0)}_{V+A}(s)
\right]_{D=2}\ .
\label{longd2}\end{equation}
The expression for $\left[\Delta\Pi^{(0)}_{V+A}(s)\right]_{D=2}$ 
in terms of the running quark masses $m_q(Q^2)$ and running
coupling $\alpha_s(Q^2)$, is known
to $O(\alpha_s^3)$~\cite{ck93}.  It turns out
that, for all of the weights studied in the literature,
the integrated version of this series is non-converging,
even at the highest scale ($s_0=m_\tau^2$) allowed
by kinematics~\cite{kmtauprob,ckp98,pp98}.  This is true
whether the integrated series is organized using
FOPT or CIPT{\begin{footnote}{As an example of the bad convergence,
for $s_0=m_\tau^2$
and the kinematically-weighted $(0,0)$ spectral weight pFESR,
the integrated series behaves
as $\sim 1+0.99+1.24+1.59$ in FOPT and $\sim 1+0.78+0.78+0.90$
in CIPT~\cite{ALEPH99}.}\end{footnote}}.
Most of the recent analyses in the literature deal with this problem by 
nonetheless retaining the longitudinal contributions out
to $O(\alpha_s^3)$, combining them with the $(0+1)$ contributions, 
and considering the sum, truncated at $O(\alpha_s^3)$.  Attempts
are made to assign errors to this truncated sum
which are sufficiently conservative to 
take into account the bad behavior of the $(0)$ part of the 
series~\cite{ckp98,ALEPH99,pp99,DHPPC00,kkp00}.  Exceptions
are (1) Ref.~\cite{km00}, which makes a subtraction of the longitudinal
contributions to the spectrum based on sum rule analyses of
the strange scalar and pseudoscalar channels~\cite{kmlong01},
and then works with pFESR's for the better behaved $(0+1)$
correlator and (2) Ref.~\cite{newpichALEPH}, which truncates
the $(0,0)$ and $(1,0)$ spectral weight pFESR's at 
$O(\alpha_s^3)$, but the $(2,0)$ spectral weight pFESR
at $O(\alpha_s^2)${\begin{footnote}{For the
$(0,0)$ and $(1,0)$ spectral weights,
the series for the sum of the $(0+1)$ and $(0)$ $D=2$ contributions
is decreasing with increasing order out to $O(\alpha_s^3)$;
for the $(2,0)$ spectral weight, however, the bad
behavior of the longitudinal part wins out earlier, and the
$O(\alpha_s^3)$ term is larger than the $O(\alpha_s^2)$ term.
The $k$-dependence of the truncation scheme 
of Ref.~\cite{newpichALEPH} for the $(k,0)$
spectral weight pFESR's results from the ansatz of truncating
the series at the point beyond which the terms begin to increase in
size.}\end{footnote}}.  

The badly-converged $D=2$ longitudinal OPE series for 
the $(k,0)$ spectral weight pFESR's turns out to have another
problem, namely that, when
employed with either of the truncation schemes noted 
above, it produces an unphysical decrease with $k$
in the extracted value of $m_s$~\cite{kmlong01}.  For such analyses
there is thus an additional theoretical systematic error 
not accounted for in the errors quoted in the literature.

It is easy to see how the problem arises.  
Since, apart from the pion pole, the
$ud$ $(J)=(0)$ spectral function is negligible, the non-pole
(``continuum'') part of the spectral function of $\Delta\Pi^{(0)}_{V+A}$,
$\Delta\rho^{(0)}_{V+A}$, is negative definite.  Thus, in
the $(k,0)$ spectral weight pFESR, where it occurs weighted
by $-2y_\tau\left( 1-y_\tau\right)^{2+k}$, it yields
a strictly positive contribution to the integrand over the whole of the
continuum region $s>s_{th}=(m_K+m_\pi)^2$.  Since 
$0<1-y_\tau <1-s_{th}/m_\tau^2=0.87$ in this region, the integrated
continuum contribution, $\left[\Delta^{(k,0)}\right]_L^c$,
is {\it necessarily} a decreasing
function of $k$, and must satisfy the
rigorous inequality
\begin{equation}
\left[\Delta^{(k+1,0)}\right]_L^c< 0.87 \left[\Delta^{(k,0)}\right]_L^c\ .
\label{rigorous}
\end{equation}
The sum of the longitudinal pion and kaon pole contributions is also
positive, and a (slowly) decreasing function of $k$.

The bound in Eq.~(\ref{rigorous}),
though rigorous, is overly conservative.
One would, in fact, expect the continuum contribution to be
dominated by the $K_0^*(1430)$ and $K(1460)$ resonances.  Since
the masses and widths of the two resonances happen to be comparable,
one can make an improved estimate for the relation of the
$\left[\Delta^{(k,0)}\right]_L^c$ for different $k$
by integrating over a Breit-Wigner profile with the average mass
and width.  The result is that, in the limit of resonance
dominance of the continuum longitudinal contributions, one
would expect~\cite{kmlong01}
\begin{eqnarray}
&&\left[\Delta^{(1,0)}\right]_L^c\simeq 0.44 \left[\Delta^{(0,0)}\right]_L^c
\nonumber\\
&&\left[\Delta^{(2,0)}\right]_L^c\simeq 
0.22 \left[\Delta^{(0,0)}\right]_L^c\ .
\label{resonancedomination}
\end{eqnarray}

We can contrast these physical constraints with what is implied by
the truncated OPE representation.  To do so we consider the 
integrated $(k,0)$ OPE representations (using
the combined fit $m_s$ central value of Ref.~\cite{pp99} as input,
to be specific) and subtract from them the very accurately known
pion and kaon pole spectral terms.  This leaves the implicit
OPE representation for the longitudinal continuum contributions.
The resulting longitudinal contributions turn out to be
comparable to, or larger than, the corresponding 
longitudinal pole contributions 
for all the cases under consideration ($k=0,1,2$).
The longitudinal contributions are also larger than the 
$(0+1)$ contributions for all cases.
With the uniform-in-$k$ truncation scheme, one finds that the
continuum longitudinal contributions predicted by
the OPE are in the ratios $1:1.16:1.42$
for the $(0,0)$, $(1,0)$ and $(2,0)$ cases, respectively.
This fails to satisfy even the weak rigorous inequalities implied
by Eq.~(\ref{rigorous}), let alone the expectations
based on resonance dominance given in Eqs.~(\ref{resonancedomination}).
If one instead uses the $k$-dependent truncation scheme of
Ref.~\cite{newpichALEPH}, the truncated OPE implies continuum
longitudinal contributions in the ratios
$1:1.16:0.83$, still rather far from those implied by the
physical constraints.  

The failure of the OPE to satisfy the physical
constraints (which must necessarily be reflected in the spectral integrals)
means that the $m_s$ values extracted in the
separate $(k,0)$ spectral weight analyses will themselves
have an unphysical $k$-dependence.  Since, for fixed $m_s$, the higher $k$
OPE contributions are too large relative to the $(0,0)$ contributions, 
successively smaller values of $m_s$
will be required to produce a match between the OPE and
spectral integrals as $k$ is increased.
This trend is seen clearly in the results
of the uniform-in-$k$ truncation scheme
for the $(k,0)$ analyses reported in the literature, and
in Table 2 above.
The size of the resulting systematic uncertainty on
$m_s(2\ {\rm GeV})$ is likely to be at least 
$20-25\ {\rm MeV}$ for the $(1,0)$ and $(2,0)$ analyses~\cite{kmlong01}.

In order to avoid the problems with the OPE representation
of the longitudinal contributions, one can work with pFESR's
for the $(0+1)$ correlator difference alone.  Since the $us$
spin separation above $1\ {\rm GeV}$ is not available, however,
some external input is required in order to allow a determination of
the longitudinal continuum subtraction to be performed
in converting from the measured to the purely $(0+1)$
spectral distribution.  

One possibility, tried
in Ref.~\cite{ALEPH99}, is to identify experimentally the $K_0^*(1430)$ and
$K(1460)$ decay modes (expected to dominate the 
non-pole part of the $(J)=(0)$ spectral integral), and
subtract these contributions, plus the longitudinal
pole contributions, from the measured spectrum.  Since
neither the $K_0^*(1430)$ nor $K(1460)$ decays of the $\tau$
have been detected to date, this is not yet practical, but
may well become so with the new B factory data.  

It is
also possible to compute the continuum $(J)=(0)$ 
subtraction if one knows the $K_0^*(1430)$ and $K(1460)$
decay constants~\cite{kmlong01}.  These can be estimated
with $\sim 10-20\%$ accuracy using sum rules for the correlators
involving the divergences of the $us$ vector and axial vector
currents~\cite{kmlong01,kmps01}.  An alternate possibility
for the strange scalar continuum contributions is to use
analyticity, unitarity and the existence of an Omnes relation
for the timelike scalar $K\pi$ form factor to compute
the strange scalar spectral function.  This was done, ignoring
the effects of channel coupling above the $K_0^*(1430)$,
in Ref.~\cite{cfnp}.  An improved version of this analysis,
which includes the effects of channel coupling using a model constrained
by ChPT and known short-distance physics, has also recently
been performed~\cite{jop}.  The effect of channel coupling on
the spectrum in the $K_0^*(1430)$ region is found to be small.
The $K_0^*(1430)$ decay constants found in the sum rule
analysis of Refs.~\cite{kmlong01,kmps01} and the coupled
channel analysis of Ref.~\cite{jop} are in good agreement
within errors, suggesting that the uncertainties in the theoretical
determination of the longitudinal subtraction are under
control, certainly at the $\sim 20-40\%$ level.
Since, for practical reasons, the weights employed in current pFESR 
analyses necessarily strongly suppress contributions from the region
above $1\ {\rm GeV}^2$, where $us$ spectral errors are large,
such a level of uncertainty on the longitudinal subtraction
produces an uncertainty in the extracted
value of $m_s$ which is negligible compared to the other sources of
error~\cite{km00,gmdpf00,kmlong01}.

An analysis of the $(0+1)$ type was performed in Ref.~\cite{km00}.
The analysis employs three weights, constructed
so as to (1) strongly suppress $us$ contributions from the region above
$1\ {\rm GeV}^2$, (2) improve the convergence of the integrated
$(0+1)$ OPE $D=2$ series, (3) reduce to some extent the strong
$ud$-$us$ cancellation present for the kinematically-weighted
$(0+1)$ spectral integral, and (4) suppress, as much as possible,
potential contributions from unknown higher dimension ($D>6$)
condensates.  The second point is of relevance because the known terms
of the $D=2$ series for $\Delta\Pi^{(0+1)}_{V+A}$ suggest
potentially slow convergence{\begin{footnote}{In 
estimating the $D=2$ truncation errors it is 
important to bear in mind that,
for the $(k,0)$ spectral weight cases, there is an accidental suppression
of the second order CIPT contour integral caused by cancellations
between contributions from different parts of the contour.  This
cancellation does not persist to higher orders, so taking the
second order term as an estimate of the truncation error
is likely to produce a significant underestimate of this
uncertainty.}\end{footnote}}.  
Explicitly,
in the $\overline{MS}$ scheme~\cite{ck93},
\begin{eqnarray}
&&\left[ \Delta\Pi^{(0+1)}_{V+A}(Q^2)\right]_{D=2}=
{\frac{3m^2_s(Q^2)}{2\pi^2Q^2}}\left[ 1+{\frac{7}{3}} a\right. \nonumber \\
&&\left. \qquad +19.9332 a^2+\cdots \right] \ .
\label{d2}\end{eqnarray}
Both the running coupling and
running mass are known to 4-loop order~\cite{beta4,gamma4}.
Details of the treatment of higher $D$ contributions may be found in
Ref.~\cite{km00}.  Regarding the fourth point,
note that, for the higher degree spectral weights
employed in the literature,  the modified
transverse kinematic weights, 
\begin{equation}
w_{(0+1)}^{(k,n)}(y)
=y^n\, \left( 1-y\right)^{2+k}\, \left( 1+2y\right)\ ,
\label{spwt01}\end{equation}
involve uncomfortably large coefficients of the $y^m$ terms
with $m>2$, which correspond to integrated
OPE contributions proportional to condensate combinations with $D=2m+2>6$.
This includes the $(1,0)$ and $(2,0)$ spectral weights,
whose pFESR's are favorable
from the point of view of reduced $ud$-$us$ cancellation.
These pFESR's are thus unfavorable from the point of view of possible
unknown higher dimension OPE contributions.  The weights of Ref.~\cite{km00}
have been designed to avoid large coefficients for this reason.

Especially after the most recent update of $R_{us}$, the $ud$-$us$
cancellation creates sizeable errors for the weights $w_{10}$
and $\hat{w}_{10}$ of Ref.~\cite{km00}.  The remaining
weight, $w_{20}(y)$, whose form is given explicitly in Ref.~\cite{km00},
has a profile in the spectral integral region intermediate between that of 
$w_{(0+1)}^{(1,0)}(y)$ and $w_{(0+1)}^{(2,0)}(y)$.  As such, it
has a lesser degree of $ud$-$us$ cancellation than the former,
but somewhat greater degree of cancellation than the latter.
Whereas $w_{(0+1)}^{(2,0)}(y)=1-2y-2y^2+8y^3-7y^4+2y^5$
has two uncomfortably large coefficients, $8$ and $-7$, however,
the largest coefficient in $w_{20}$, associated with
$D=8$ contributions, is $2.1$.  Updating this analysis for
the new $R_{us}$ value (following the procedure described in 
Ref.~\cite{DHPPC00} for the modification of the
$us$ spectral distribution), truncating the $D=2$ OPE
series at $O(a^2)$, and representing the $D>6$ contributions
by an effective $D=8$ term, fit to data, we find
\begin{equation}
m_s(2\ {\rm GeV})=123\pm 18\pm 15\ {\rm MeV}\ ,
\label{w20ckmu}\end{equation}
for CKMU input and
\begin{equation}
m_s(2\ {\rm GeV})=104\pm 18\pm 17\ {\rm MeV}\ ,
\label{w20ckmn}\end{equation}
for CKMN input.  The first error in each
case is experimental, the second
theoretical.  The theoretical error
is dominated by the estimate of the error associated with truncating
the $D=2$ series.
The size of the data errors,
and the quality of the match between the OPE and hadronic
sides of the $w_{20}$ FESR which results, are shown in
Figure 4.

Possibilities for improving the $(0+1)$ determination 
of $m_s$ will be described in the final section.

\begin{figure}[h]
\unitlength1cm
\caption{Optimized OPE fit to the spectral integrals for the $w_{20}$ pFESR.
The $(D=4)$ and $D=6$ OPE contributions have been
subtracted from the data integrals.  The solid line corresponds to
the best fit for $m_s$ and the coefficient for
the $D_{eff}=8$ effective operator used to represent
$D>6$ effects.  The data integrals and
errors correspond to the ALEPH data and covariance matrix.}
\begin{minipage}[t]{8.cm}
\begin{picture}(7.8,7.8)
\epsfig{figure=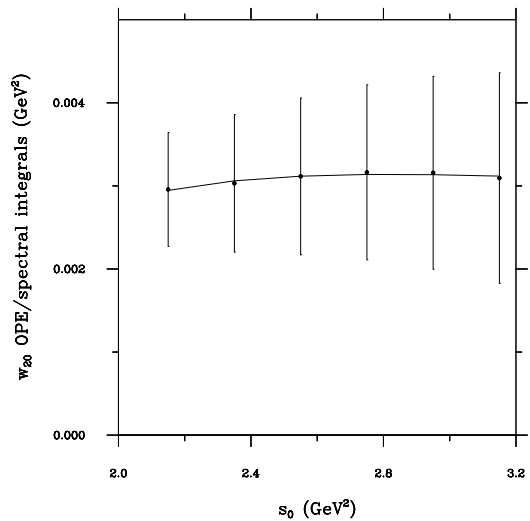,height=7.6cm,width=7.6cm}
\end{picture}
\end{minipage}
\label{Figure:4}\end{figure}

\section{CHIRAL LIMIT VALUES OF THE $K\rightarrow\pi\pi$ 
EWP MATRIX ELEMENTS}
In the Standard Model, the effective strangeness-changing non-leptonic decay
Hamiltonian takes the form
\begin{eqnarray}
&&H_{w,eff}^{\Delta S=1}={\frac{G_F}{\sqrt{2}}}V_{ud}V^*_{us}
\sum_{k=1}^{10} \left[ z_k +\tau y_k\right] Q_k\ ,
\label{heff}
\end{eqnarray}
where $\tau\equiv -V_{td}V^*_{ts}/V_{ud}V^*_{us}$ and
the explicit forms of the effective operators
$Q_1,\cdots ,Q_{10}$ may be found in Ref.~\cite{burasrev}.  The
Wilson coefficients, $z_k$ and $y_k$, are known
to NLO in $\alpha_s$~\cite{burasrev,buraswc,martinelliwc}.  
The task of determining
physical weak decay amplitudes, such as those relevant
to $\epsilon^\prime /\epsilon$, is thus reduced to the problem of
evaluating the relevant low-energy matrix elements of the
operators $Q_k$.

Since CP violation in the Standard Model requires the participation
of all three generations, $\epsilon^\prime /\epsilon$
receives contributions only from the gluonic penguin operators
$Q_{3-6}$ and EWP operators
$Q_{7-10}$.  For the large physical top quark mass, it turns out
that the dominant contributions are those
asociated with $Q_6$ and $Q_8$~\cite{burasrev}.
In the $\overline{MS}$ NDR scheme, for example~\cite{burasrev},
\begin{eqnarray}
&&{\frac{\epsilon^\prime}{\epsilon}} = 20 \times 10^{-4}
\left( {\frac{{\rm Im} \lambda_t}{1.3 \cdot 10^{-3}}}\right)
\left[ - 2.0\ {\rm GeV}^{-3} \right. \nonumber \\
&&\left.\quad \times \langle (\pi\pi)_{I=0} |
Q_6 | K^0\rangle_{2\ {\rm GeV}}
(1 - \Omega_{\rm IB}) \right. \nonumber \\
&&\left. \quad - 0.50\ {\rm GeV}^{-3} \cdot
\langle (\pi\pi)_{I=2} | Q_8 | K^0\rangle_{2~{\rm GeV}}
\right. \nonumber\\
&&\left.\quad - 0.06 \right] \ \ ,
\label{r1}
\end{eqnarray}
where $\Omega_{IB}$ is an isospin-breaking correction~\cite{strongIB}
and $\lambda_t=V_{td}V^*_{ts}$.

By good fortune it turns out to be
possible to use hadronic $\tau$ decay data to determine the values
of the $K\rightarrow\pi\pi$ EWP matrix elements 
$\langle \pi\pi\vert Q_{7,8}\vert K\rangle$ in the chiral
limit.  The $Q_8$ matrix element is a crucial ingredient
in our understanding of the Standard Model prediction
for $\epsilon^\prime /\epsilon$.  Although the $Q_7$ matrix
is not of equivalent phenomenological interest,
it is likely to be of relevance as a test of lattice
techniques for the evaluation of weak matrix elements,
particularly the reliability of the extrapolations
to physical light quark masses.  We briefly review
the fortuitous circumstances which make the evaluation
of these matrix elements possible.

From the explicit form of $Q_7$, $Q_8$,
\begin{equation}
Q_7={\frac{3}{2}}{\bar{s}}_a\gamma_\mu \left( 1-\gamma_5\right) d_a
\sum_q e_q{\bar{q}}_b\gamma^\mu\left( 1-\gamma_5\right) q_b
\nonumber\end{equation}
and
\begin{equation}
Q_8={\frac{3}{2}}{\bar{s}}_a\gamma_\mu \left( 1-\gamma_5\right) d_b
\sum_q e_q{\bar{q}}_b\gamma^\mu\left( 1-\gamma_5\right) q_a
\ ,\nonumber
\end{equation}
where $a,b$ are color labels, one sees that both transform
as $8_L\times 8_R$ under $SU(3)_L\times SU(3)_R$.  As a result,
the $K\rightarrow\pi\pi$ matrix elements survive in the chiral
limit (in contrast to the gluonic penguin matrix elements)
and hence can be evaluated using soft pion/kaon techniques.
The results, for the $I=2$ final $\pi\pi$ state, are~\cite{dgsoftpi}
\begin{equation}
\langle (\pi\pi )_2 | {\cal Q}_8 | K^0\rangle_\mu  = -
{\frac{2}{3F_0^{3}}}~\left[ \langle O_1 \rangle
+{\frac{3}{2}}\langle O_8 \rangle\right]_\mu
\end{equation}
and
\begin{equation}
\langle (\pi\pi )_2 | {\cal Q}_7|K^0\rangle_\mu = - 
{\frac{2}{F_0^{3}}}~\langle O_1\rangle_\mu \ \ , 
\label{r4} 
\end{equation}
where $F_0$ is the pion decay constant in the chiral limit
and the operators are given by
$O_{1,8}$ are 
\begin{eqnarray}
O_1 &\equiv& {\bar q} \gamma_\mu {\tau_3 \over 2} q
~{\bar q} \gamma^\mu {\tau_3 \over 2} q \nonumber \\
&&\quad - {\bar q} \gamma_\mu \gamma_5 {\frac{\tau_3}{2}} q
~{\bar q} \gamma^\mu \gamma_5 {\frac{\tau_3}{2}} q \ ,
\nonumber \\
O_8 &\equiv& {\bar q} \gamma_\mu \lambda^a
{\frac{\tau_3}{2}} q~{\bar q} \gamma^\mu \lambda^a {\frac{\tau_3}{2}} q
\nonumber \\
&&\quad - {\bar q} \gamma_\mu \gamma_5 \lambda^a {\frac{\tau_3}{2}} q
{\bar q} \gamma^\mu \gamma_5 \lambda^a {\frac{\tau_3}{2}} q
\ \ .
\end{eqnarray}

The connection to $\tau$ decay data arises from the fact that
precisely the same vacuum matrix elements $\langle O_{1,8}\rangle$
determine the $D=6$ part of the flavor $ud$ V-A correlator,
$\Delta\Pi_{ud}\equiv \Pi_{V;ud}^{(0+1)} - \Pi_{A;ud}^{(0+1)}$.
Explicitly~\cite{cdgm01}
\begin{equation}
\left[ \Delta\Pi_{ud}\right]_{D=6}= a_6(\mu )+b_6(\mu )\, log(Q^2/\mu^2)\ ,
\label{d6vma}\end{equation}
with
\begin{eqnarray}
a_6 (\mu )&=& 2\, \left[ \langle \left( 2\pi \alpha_s 
+A_8 \alpha_s^2\right) O_8 \rangle \right.\nonumber\\
&&\left. \qquad +A_1 \langle \alpha_s^2 O_1 \rangle \right]_\mu\ ,\nonumber \\
b_6 (\mu ) &=& 2\, \left[\alpha_s^2\left( B_8 \langle O_8 \rangle +
B_1 \langle O_1 \rangle \right)\right]_\mu \ \ ,
\label{r15a}
\end{eqnarray}
where $A_{1,8}$ and $B_{1,8}$ are coefficients depending on
the renormalization scheme, the prescription for the treatment
of $\gamma_5$, and the evanescent operator basis~\cite{burasrev}.
Values employing the same evanescent operator basis as used
for computing the Wilson coefficients of 
$H_{w,eff}$~\cite{buraswc,martinelliwc} and for both the HV and NDR
prescriptions for $\gamma_5$, may be found in Ref.~\cite{cdgm01}.
$\langle O_1\rangle$ is suppressed
relative to $\langle O_8\rangle$ in the large $N_C$ expansion.
Dispersive determinations (to be discussed below) also indicate
that it is much smaller numerically.  Thus, both the dominant
CP violating $K\rightarrow\pi\pi$ $Q_8$ matrix element and
the $D=6$ part of the $ud$ V-A correlator are essentially
determined by the same quantity, $\langle O_8\rangle$.
One can therefore construct either dispersive sum rules or
pFESR's for the V-A correlator in an attempt to extract this
matrix element using the $ud$ V-A hadronic $\tau$ decay spectral data.

Dispersive sum rules for $\langle O_{1,8}\rangle$ were first
considered in Ref.~\cite{dgsoftpi} and re-analyzed in
Ref.~\cite{cdgm01}, taking into account the $D>6$ contributions
discussed in Ref.~\cite{dghigherd} and NLO radiative
corrections.  With $\Delta\rho (s)$
the spectral function of $\Delta\Pi_{ud}$, one finds, 
for $\langle O_1\rangle$,
\begin{equation}
\langle O_1 \rangle_\mu - {\frac{3 C_8}{8 \pi}}
\langle \alpha_s O_8 \rangle_\mu
= \bar{I}_1 (\mu) 
\label{dgeq1}\end{equation}
where the scheme-dependent coefficient $C_8$ is given in
Ref.~\cite{cdgm01}, and
\begin{equation}
\bar{I}_1 (\mu) = {\frac{3}{(4 \pi)^2}} 
\left[ I_1 (\mu) + H_1 (\mu) \right] \ , 
\label{i1def}
\end{equation}
with
\begin{eqnarray}
&&I_1 (\mu) = \int_0^\infty ds\ s^2 \ln
\left({\frac{s + \mu^2}{s}} \right) ~\Delta\rho(s)\nonumber \\
&&H_1 (\mu)=
\int_{\mu^2}^\infty dQ^2 \, Q^4~\left[\Delta\Pi_{ud} (Q^2)\right]_{D>6}. 
\label{r17}  
\end{eqnarray}
Similarly, for $\langle O_8\rangle$, one has
\begin{eqnarray}
&&\langle\left( 2\pi \alpha_s +\alpha_s^2\right) O_8\rangle_\mu
+ A_1 \langle \alpha_s^2 O_1 \rangle_\mu 
\nonumber\\
&&\qquad \qquad
= 2 \pi \alpha_s (\mu)  \bar{I}_8 (\mu) 
\label{va38}
\end{eqnarray}
where
\begin{equation}
\bar{I}_8 (\mu) = \frac{1}{2 \pi \alpha_s (\mu)} 
\bigg[ I_8 (\mu) - H_8 (\mu) \bigg] \ , 
\label{i8def}
\end{equation}
with
\begin{eqnarray}
&&I_8 (\mu )=  \int_0^\infty ds\ {\frac{s^2\mu^2}{ s + \mu^2}}
~\Delta\rho (s) \nonumber\\
&&H_8 (\mu )= \mu^6 \left[\Delta\Pi_{ud} (\mu )\right]_{D>6}. 
\label{r19} 
\end{eqnarray}

Although the spectral integrals $I_{1,8}(\mu )$ extend over
the whole range $0<s<\infty$ and not just $0<s<m_\tau^2$,
it turns out to be possible to use the relations above, together with
hadronic $\tau$ decay data, to evaluate these integrals
with good accuracy.  This is made possible by the fact that the $ud$ V-A
correlator satisfies three classical chiral sum rules in the
chiral limit, the two Weinberg sum rules~\cite{weinbergsr}
and the sum rule for the chiral limit pion electromagnetic
splitting~\cite{empisr}.  These sum rules involve integrals
over the $ud$ V-A spectral function, $\Delta\rho (s)$, but with
weights $w(s)=1,s$ and $s\, log(s/\Lambda^2)$ (for any $\Lambda$), 
respectively, rather than those relevant to $\langle O_{1,8}\rangle$.

Any weighted integral of the form $\int_0^\infty ds\, w(s)\Delta\rho (s)$
can be re-written as
\begin{eqnarray}
&&\int_0^\infty ds\, \left[ c_1+c_2 s+ c_3 s\, 
log\left({\frac{s}{\Lambda^2}}\right)\right]
\nonumber\\
&&\qquad + \int_0^{\infty}ds\, \Delta w(s)\Delta\rho(s)
\label{rwm}\end{eqnarray}
where $\Delta w(s) = w(s)-c_1-c_2 s - c_3 s\, log(s/\Lambda^2)$.
The first integral in Eq.~(\ref{rwm}) is determined, for any $c_{1-3}$,
by the chiral limit values of $f_\pi$ and the
pion squared electromagnetic mass splitting, as a
consequence of the three classical chiral
sum rules. So long as it is possible to make $\Delta w(s)$ small in
the region above $s=m_\tau^2$ by an appropriate
choice of $c_{1-3}$, the original integral can be evaluated
in terms of the $\tau$ spectral data and the chiral limit
input.  The details of how this ``residual weight method'' (RWM), 
is implemented, and the procedure for minimizing the resulting
errors, may be found in Ref.~\cite{cdgm01}.  See also 
Refs.~\cite{bgp01,drp01,narison01} for other approaches
to employing dispersive sum rules, or FESR's, to evaluate the chiral limit
$Q_{7,8}$ $K\rightarrow\pi\pi$ matrix elements.

The RWM method turns out to produce errors on
the spectral integrals $I_{1,8}(\mu )$ which
increase as $\mu$ is increased.
To be able to work at lower scales, where the errors are small,
however, one has to worry about the presence of the higher
dimension $D>6$ contributions $H_{1,8}(\mu )$.  An alternate approach to
determining the $D=6$ term in the $ud$ V-A OPE is to employ
appropriately designed pFESR's.  These have the advantage
of requiring data only over a limited range $s<s_0$,
which may be chosen to lie entirely in the range
allowed by $\tau$ decay kinematics.  In addition, they can,
in principle, be employed to extract the
higher dimension coefficients in the V-A OPE
which are needed for an empirical determination 
of $H_{1,8}$.  Such a determination would then allow
for a low-scale version
of the RWM analysis.  Since, at scales $\mu\sim 2\ {\rm GeV}$
the contributions to $I_{1,8}(\mu )$ associated with
the chiral constraints are more than $50\%$ again larger than
the integrals of the residual weights over the range
of $s$ covered by the $\tau$ decay data, the pure pFESR and
low-scale RWM ``hybrid'' analyses
are largely independent, and provide
a useful self-consistency check on one another.
Details of these two analyses
may be found in Refs.~\cite{cdgm02}.  

The preliminary results
from this analysis show excellent consistency between the
two methods for the $Q_8$ matrix elements,
and are listed in Table 3.  Note that the $\langle O_1\rangle$
contribution to $a_6$ is only $\sim -3\%$ of that associated
with $\langle O_8\rangle$.  The direct pFESR extraction 
thus determines only 
$\langle O_8\rangle${\begin{footnote}{The smallness of
the $\langle O_1\rangle$ contribution is a result of
two factors: the color suppression of $\langle O_1\rangle$
relative to $\langle O_8\rangle$ and the additional 
factor of $\alpha_s$ in the coefficient of $\langle O_1\rangle$.
The results of the dispersive analysis are, in fact,
used to evaluate the tiny $\langle O_1\rangle$ contribution
to $a_6$.}\end{footnote}}.  Thus, although one may perform a hybrid
analysis for $\langle O_1\rangle$, no pure pFESR analysis is
possible.  An independent test of the hybrid extraction of 
$\langle O_1\rangle$
can, however, be obtained by performing a RWM analysis for
$\langle O_1\rangle$ at a scale sufficiently high that
$H_1(\mu )$ may be taken to be zero (say $\mu\sim 4\ {\rm GeV}$)
and then evolving the results down to $\mu =2\ {\rm GeV}$
using the known anomalous dimension
matrix~\cite{burasanomalous}.
The results of this test are shown in Table 4.

The results of the pFESR analysis correspond
to an EWP contribution to $\epsilon^\prime /\epsilon$, in the
chiral limit, of
\begin{equation}
\left[ \epsilon^\prime /\epsilon\right]^{\chi L}_{EWP}
=\left( -16.2\pm 3.4\right)\times 10^{-4}\ ,
\label{ewpepe}\end{equation}
and to deviations from the vacuum saturation value
for the $Q_8$ matrix element $B_8=1.7$ and $1.9$ for
the NDR and HV schemes, respectively.  These are a factor
of $\sim 2$ larger than the values produced by most models 
employed previously in the literature.  The result of
Eq.~(\ref{ewpepe}) is similar in magnitude, but opposite in sign,
to the current experimental determination
\begin{equation}
\left[ \epsilon^\prime /\epsilon\right]_{exp}
=\left( 18\pm 4\right)\times 10^{-4}\ .
\end{equation}
A rapid change in $\epsilon^\prime /\epsilon$ as $m_s$ is
varied from $0$ to its physical value
is thus required if the Standard Model is to explain
the experimental result.  

\begin{table*}
\begin{center}
\caption{pFESR and ``hybrid'' results for the matrix elements
$M_{7,8}\equiv 
\langle (\pi\pi )_{I=2}\vert Q_{7,8}\vert K^0\rangle_{2\ {\rm GeV}}$.
The hybrid analysis is an RWM analysis at $\mu =2\ {\rm GeV}$,
using pFESR input to evaluate $H_{1,8}(2\ {\rm GeV})$. Results are
in units of GeV$^6$.}
\label{table:3}
\newcommand{\m}{\hphantom{$-$}}
\newcommand{\cc}[1]{\multicolumn{1}{c}{#1}}
\renewcommand{\tabcolsep}{2pc} 
\renewcommand{\arraystretch}{1.2} 
\vskip .15in\noindent
\begin{tabular}{@{}llll}
\hline
Scheme&Method&$M_8(2\ {\rm GeV})$&
$M_7(2\ {\rm GeV})$\\
\hline
NDR&hybrid&$1.65\pm 0.45$&$0.21\pm 0.03$\\
&pFESR&$1.62\pm 0.34$&$-$\\
HV&hybrid&$1.84\pm 0.46$&$0.46\pm 0.08$\\
&pFESR&$1.80\pm 0.36$&$-$\\
\hline
\end{tabular}\\[2pt]
\end{center}
\end{table*}

\begin{table*}
\begin{center}
\caption{Comparison of the ``hybrid'' determination of
$\langle O_1(2\ {\rm GeV})\rangle$ and the results
obtained by evolving the $\mu=4\ {\rm GeV}$ scale
RWM results to $\mu =2\ {\rm GeV}$ using the known
anomalous dimension matrix.  Results are in units of
$10^{-4} \ {\rm GeV}^3$.}
\label{table:4}
\newcommand{\m}{\hphantom{$-$}}
\newcommand{\cc}[1]{\multicolumn{1}{c}{#1}}
\renewcommand{\tabcolsep}{2pc} 
\renewcommand{\arraystretch}{1.2} 
\vskip .15in\noindent
\begin{tabular}{@{}lcc}
\hline
Scheme&Hybrid Analysis&Evolution from $\mu =4\ {\rm GeV}$\\
\hline
NDR&$-(0.70\pm 0.11)$&$-(0.53\pm 0.34)$\\
HV&$-(1.52\pm 0.27)$&$-(1.64\pm 0.18)$\\
\hline
\end{tabular}\\[2pt]
\end{center}
\end{table*}

\section{COMMENTS/PROSPECTS}
The hadronic $\tau$ decay data base will increase significantly
with the data from CLEO-C, the new B factory experiments
and (eventually) the Beijing $\tau$-charm factory.  In this section
the question of what improvements in the determinations discussed
above are likely to be made possible by this new data is
briefly discussed.

Regarding $\alpha_s$, it appears unlikely that significant
further improvement can be made since the
dominant error, even with the existing data, is theoretical,
associated with the truncation of the $D=0$ series~\cite{ALEPH99,OPALlight}.
This shows up in the direct estimates of the truncation
error based on variations in the unknown $O(a^4)$ term in
the $D=0$ part of the Adler function and/or the
renormalization scale, and also in the
deviations between the solutions for $\alpha_s$
obtained using different methods of handling the truncated
$D=0$ series{\begin{footnote}{The CIPT scheme is based on
the truncation of the Adler function at fixed order,
followed by integration around the contour $\vert s\vert =s_0$, the FOPT
scheme on the expansion of the integrated Adler function
to fixed order in $\alpha_s(s_0)$.  Since the integral of
$\left[\alpha_s(Q^2)\right]^k$ can be written as a series
in $\alpha_s(s_0)$ beginning at order $k$, the two schemes
differ by terms higher order in $\alpha_s(s_0)$.
The difference between the results $\alpha_s(m_\tau )=0.345$
and $\alpha_s(m_\tau )=0.326$, obtained from the CIPT and FOPT
treatments of the same pFESR, at the same value of $s_0$,
thus provides one estimate for the size of the truncation
error.}\end{footnote}}.

The situation is quite different for $m_s$, where significant
progress can be expected in the near future.  The B factory
data should drastically reduce the errors on the $us$ spectral
distribution.  In addition to the obvious advantage
of producing a reduction in the size of the errors 
on $m_s$ for existing analyses, such improvements
should also make possible the use of alternate sum rules
not practically useful at present because of the size 
of the spectral integral errors when evaluated with current data.

This latter point may be relevant to improving
less-than-optimal features of existing analyses.  An
example is the current sensitivity to $\vert V_{us}\vert$.  
At present, the large $us$ experimental errors above the $K^*$ force one 
to work with weights which strongly suppress
the high-$s$ part of the spectrum.
These weights turn out to produce a rather high level of $ud$-$us$ 
cancellation, and hence a significant sensitivity to
$\vert V_{us}\vert$.  It is likely that, with improved data,
the reduction in the level of high-$s$ suppressions required
will allow the construction of alternate weights with a reduced
level of $ud$-$us$ cancellation.

An important point to be stressed regarding the extraction of $m_s$ 
is that, given the sizeable theoretical systematic
problems associated with the 
behavior of the OPE representation of the longitudinal
$us$ contributions, the inclusive spectral weight analysis 
method should almost certainly be abandoned.  There are two
ways to deal with the longitudinal subtraction that must
be made if one wishes pursue sum rules based on
the $(0+1)$ correlator.  

One is the theoretical approach outlined
above.  Note that the continuum longitudinal subtraction 
does not need to be known to very high accuracy since it
actually has little impact on the $(0+1)$ sum rules, at least
for the weights studied so far in the literature.  For example,
using the $K(1460)$ and $K_0^*(1430)$ decay constants
extracted from the sum rule analyses of the scalar and
pseudoscalar channels~\cite{kmps01} to make the subtraction,
one finds that the shift in the values of
the $(0+1)$ spectral integrals induced by the subtraction is less than
$4\%$ for $s_0$ in the range $s_0>2\ {\rm GeV}^2$,
even for the least rapidly falling of the $(0+1)$ weights
employed in the literature (the $(0,0)$ spectral weight).
Thus even a $100\%$ uncertainty in the longitudinal
subtraction corresponds to an error vastly smaller
than that associated with the use of the badly-behaved
longitudinal OPE representation.  

The second approach is experimental, and
likely to be feasible, at least to some extent, with
the new B factory data.  If
current theoretical estimates for the $K_0^*(1430)$ decay 
constant~\cite{kmps01,jop} are correct, the 
$\tau\rightarrow\nu_\tau K_0^*(1430)$ branching fraction
should be $\sim 6\times 10^{-5}$ (a factor of $\sim 8$ 
smaller than the current experimental
upper bound), a level which may be reachable in the new
B factory experiments.  Since the
$K_0^*(1430)$ decays essentially entirely to $K\pi$, the
only experimental complication (apart from rate) is
the $\sim 7\%$ $K\pi$ branch of the $K^*(1410)$.  Based on
the central value of the ALEPH determination,
$B(\tau\rightarrow K^*(1410)\nu_\tau )=\left( 1.5{+1.4\atop -1.0}
\right)\times 10^{-3}$, one would expect
a $K\pi$ background in the $K_0^*(1430)$
region at the $\sim 1\times 10^{-4}$ level, and with a
width somewhat smaller than that of the $K_0^*(1430)$.  
The $\tau\rightarrow\nu_\tau K(1460)$ mode appears less amenable 
to experimental identification;
confirmation of the theoretical
estimates for the $K_0^*(1430)$ decay constant would, however,
serve as strong evidence in favor of the corresponding 
$K(1460)$ decay constant estimate.  Even an upper bound
on $B(\tau\rightarrow K_0^*(1430)\nu_\tau )$ at the
$\sim 1\times 10^{-4}$ level would be useful since the impact of the 
longitudinal subtraction on the $(0+1)$ $m_s$ extraction,
at the predicted level for the two decay constants,
is already known to be small.
It is thus likely that the uncertainty in the impact
of the longitudinal subtraction can be made rather small {\it experimentally}
in the near future.

Regarding the $K\rightarrow\pi\pi$ matrix elements of the EWP
operators, the main improvement to be hoped for is reduced
errors in the $ud$ V-A spectral function in the region above
$2\ {\rm GeV}^2$.  This would require an improved V/A
separation of the contributions to states containing a
$K\bar{K}$ pair.  Such an improvement might be possible
with the enhanced statistics expected from the B factory
experiments, at least for quasi-two-body modes.
A reduction of the errors on the V-A spectral function would
allow one not only to reduce the errors on the extracted
EWP matrix elements, but also to sharpen tests for the absence 
of duality violation.
Recall that both the $ud$ V and A correlators display, in
general, significant duality violation for $s_0<m_\tau^2$.
Since the V and A duality violations largely cancel in the
V+A sum, the intrinsic level of duality violation in the
V-A correlator must be expected to be large.  The analysis
of Ref.~\cite{cdgm02} tests for the presence of such
violations in the pFESR's employed,
and finds none, within experimental errors, for $s_0$
above $\sim 1.8\ {\rm GeV}^2$.  The larger
errors in the upper part of the $s_0$ window, however, leave 
room for improvement in these tests, particularly in the
case of the pFESR's employed to extract the $D>6$
contributions needed as input to the hybrid RWM/pFESR version
of the analysis.

I would like conclude by acknowledging useful
exchanges with Andreas H\"ocker and Shaomin Chen
concerning the ALEPH data tabulation, and especially
the work of my collaborators on some of the topics
reported here (Joachim Kambor on
the light quark masses and related issues, and
Vincenzo Cirigliano, John Donoghue and Gene Golowich
of the $K\rightarrow\pi\pi$ matrix elements of
the EWP operators).

\end{document}